\magnification=1200
\baselineskip=16pt
\vskip 0cm
\centerline{\bf Interface Fluctuations on a Hierarchical Lattice}
\vskip 0.5cm
\centerline{Ferenc Igl\'oi}
\bigskip
\centerline{Research Institute for Solid State Physics, 
H-1525 Budapest, P.O.Box 49, Hungary}
\smallskip
\centerline {and}
\smallskip
\centerline{Laboratoire de Physique du Solide, Universit\'e Henri Poincar\'e (Nancy I), B.P. 239,}
\centerline{F-54506 Vand\oe uvre l\`es Nancy Cedex, France}
\vskip 0.5cm
\centerline{and}
\vskip 0.5cm
\centerline{Ferenc Szalma}
\bigskip
\centerline{Institute for Theoretical Physics, Szeged University}
\centerline{H-6720 Szeged, Aradi V. tere 1, Hungary}
\vskip 1cm
{\bf Abstract:}
We consider interface fluctuations on a two-dimensional layered lattice where
the couplings follow a hierarchical sequence. This problem
is equivalent to the diffusion process of a quantum particle in the presence of
a one-dimensional hierarchical potential. According
to a modified Harris criterion this type of perturbation is relevant and one
expects anomalous fluctuating behavior. By transfer-matrix techniques and by
an exact renormalization group transformation we have obtained analytical
results for the interface fluctuation exponents, which are discontinuous
at the homogeneous lattice limit.
\smallskip
PACS-numbers: 05.40.+j, 64.60.Ak, 68.35.Rh
\vfill
\eject
{\bf I. Introduction}
\vskip 1cm
Recently there is a growing interest in such natural and artificial systems
which are organized in a hierarchical way. Examples can be found in economical
organizations[1] and stock-market exchanges[2], in geological processes before
major earthquakes[3], in studies of relaxation phenomena of proteins[4],
spin glasses[5] and computer architectures[6]. Theoretically much effort have
been devoted to understand the linear dynamics (i.e. the diffusion process)
in a system with hierarchically organized energy barriers. According to
numerical[7,8] and exact[9,10] results the diffusion in such systems can be
anomalous (which is often called as "ultradiffusion"[7]), furthermore in
several models there is a dynamical phase transition[8] separating regions
with normal and anomalous diffusion. For a comprehensive review on the subject
see Ref[11].

Another subject of theoretical interest is the properties of (static) phase
transitions on hierarchical lattices. For these and other non-periodic
(quasi-periodic or more generally aperiodic) systems a relevance-irrelevance
criterion has recently been proposed[12] on the analogy of the Harris
criterion[13]
for random magnets. The cross-over exponent corresponding to a non-periodic
perturbation is given by:
$$\Phi=1+\nu D (\Omega -1) \eqno(1)$$
in terms of the $\nu$ correlation length exponent of the unperturbed system
and the wandering exponent
of the sequence $\Omega$[14]. Here $D$ denotes the number of coordinates on
which the couplings depend, c.f. for a layered system $D=1$. The perturbation
is then expected to be relevant
(irrelevant) if $\Phi>0$ ($\Phi<0$), which was indeed found in a series of
exact studies on two-dimensional layered Ising models[15,16].
For marginal sequences,
where $\Phi=0$, continuously varying critical exponents
and anisotropic scaling behavior was observed[17]. 

As far as the critical behavior on hierarchical lattices is concerned mainly
the two-dimensional layered Ising model with a one-dimensional
Huberman-Kerszberg (HK) sequence[7] and the corresponding Ising quantum chain
were studied. In numerical[18] and exact[19,20] calculations non-universal critical
behavior were found in accordance with the vanishing cross-over exponent
in eq(1), which follows from the fact that the fluctuation 
exponent of the HK sequence is $\Omega=0$[20].

In this paper we consider the interface fluctuation problem on a
layered lattice, where the couplings between the layers follow the HK
hierarchical sequence. As far as interface wandering on non-periodic lattices
are concerned the work by Henley and Lipowsky[21] has to be mentioned, who
considered
the interface roughening in two-dimensional quasicrystals.
On a layered lattice with Fibonacci-type quasi-periodicity
non-universal interface fluctuations were observed, with a continuously
varying interface
wandering exponent. This behavior is again
in accord with the relevance-irrelevance criterion, since with $\Omega=-1$
and $\nu=\nu_{\perp}=1/2$ the cross-over exponent in eq(1) is $\Phi=0$.
In our problem, on the HK lattice $\Omega=0$, thus $\Phi=1/2>0$ and the
perturbation is relevant. Therefore one expects anomalous interface fluctuations
on this lattice.

The structure of the paper is the following. We define the model in Sec.II.
The results of the transfer matrix calculations and that of an exact
renormalization group (RG) transformation are presented in Sections III. and
IV., respectively. The results are discussed in the final section.
 
\bigskip
{\bf II. Formalism}
\vskip 1cm
We consider a diagonally layered ferromagnetic spin model (c.f. the Ising model)
on the square lattice with hierarchically organized interactions. The couplings
in the $h$-th diagonal $K_h=J_h/k_B T$ are selected
from a set $(\kappa_0,\kappa_1,\kappa_2,...)$ and  $\kappa_n=n \kappa_0$,
such that
$$K_h=\kappa_n~~~,~~~h=2^n (2m+1) . \eqno(2)$$
This type of structure of the couplings (Fig.1), which shows the typical
features of ultrametric topology[5] 
was introduced by Huberman and Kerszberg[7] following the work in Ref[1].

The boundary spins on the (1,1) surfaces are fixed
in different orientations (Fig.1) and we are interested in the fluctuations
of the interface separating the (+) and
(-) regions. The interface is considered
as a continuous structurless string and complicated interface
configurations, such as overhangs and bubbles are omitted. It is generally accepted
that
to study interfacial fluctuations it is sufficient to keep only Solid-on-Solid
(SOS) type interface configurations. In this so called SOS model
the interface is geometrically represented by a directed walk or
polymer[22].

In the SOS model the interface
is characterised by its $h(x)$ height at site $x$ and the interfacial energy
is specified by the Hamiltonian:
$$-H/k_B T=\sum_x 2 K_{h(x)}~~~, \eqno(3)$$
where surface effects are omitted.
The thermodynamic properties of the interface are conveniently studied in the
transfer matrix formalism[23,24]. For our model the transfer matrix in the 
$x$-direction, parallel with the boundaries, is given by:
$$T_{h,l}=\delta_{h,l-1}e^{-2K_h}+\delta_{h,l+1}e^{-2K_l}~~~. \eqno(4)$$
Here according to eq(2) the matrix-elements are from a set
$(\epsilon_0,\epsilon_1, \epsilon_2,
... )$ and the ratio of successive terms is constant: $\epsilon_{n+1}/
\epsilon_n=R<1$. For the homogeneous system $R=1$, whereas for hierarchical
lattices $R$ measures the strength of inhomogeneity. The interface is not
likely to visit sites with a matrix-element $\epsilon_n$, $n \gg 1$, since
the corresponding probability is weighted by a factor of $R^n$.

The interfacial free energy $\sigma$ and the longitudinal
correlation length $\xi_{\parallel}$, which is measured parallel with the
boundaries, are given in terms of the leading and the
next-to leading eigenvalues of the transfer matrix $\lambda_0$ and $\lambda_1$
as:
$$\sigma=-\log \lambda_0  \eqno(5)$$
and
$$\xi_{\parallel}^{-1}=\log(\lambda_0/\lambda_1)~~~. \eqno(6)$$
The fluctuations of the interface grow on a power law scale:
$$<[h(0)-h(x)]^2> \sim x^{2w}~~~, \eqno(7)$$
where $w$ is the wandering or fluctuation exponent, which is $w=1/2$ for
homogeneous two-dimensional systems[22]. 

Another quantity of interest is the probability $P_0(x)$
that the interface after $x$-steps has the same position, i.e. $h(0)=h(x)$.
For a
walk or diffusion problem, where $x$ plays the role of the time,
$P_0(x)$ is the
autocorrelation function, which has the asymptotic behavior $P_0(x) \sim
x^{-\gamma}$. For homogeneous two-dimensional lattices $\gamma=1/2$ and
generally $w=\gamma$[8].
It could be shown by slightly modifying the derivation in Ref.[8] that the
autocorrelation function
averaged over the starting positions of the interface can be expressed
through the spectrum of the transfer matrix as:
$$\bar{P_0}(x)={1 \over L} \sum_i \left( {\lambda_i \over \lambda_0} \right)
^x=\int_{-\infty}^{1} g(\lambda) \left( {\lambda \over \lambda_0} \right)
^x d \lambda \eqno(8)$$
where $g(\lambda)=1/L \sum_i \delta(\lambda-\lambda_i)$ is the density of
states and $L$ denotes the width of the system in the $h$ direction, thus
it is the dimension of the transfer matrix.

The eigenvalues of the transfer matrix are dense at the top of the spectrum and
one can develop a scaling theory in terms of these critical eigenvalues. We
consider a critical level $\lambda_i$ of
a system with a finite width $L$ and denote by $\Delta \lambda_i=
\lambda_0-\lambda_i$ its difference from the top of the spectrum.
Changing lengths by
a factor of b=2, i.e. with $L'=L/2$ the $i$-th eigenvalue will be $\lambda_i'$,
and the difference $(\Delta \lambda_i)'$ will
scale with a factor of $b^{y_{\lambda}}$, thus
$$(\Delta \lambda_i)'=2^{y_{\lambda}} \Delta \lambda_i~~~, \eqno(9)$$
where $y_{\lambda}$ is the gap exponent. We stress that the
statement in eq(9), that all critical levels scale with the
same factor is a scaling hypothesis, which will be verified by actual
calculations in the following sections.

Using eq(9) the transformation law for the density of states is given by
$$g(\Delta \lambda)=2^{y_{\lambda}-1} g'[(\Delta \lambda)']  \eqno(10)$$
which is compatible with a power law dependence of the density of states at
the top of the spectrum:
$$g(\Delta \lambda) \sim (\Delta \lambda)^{1/y_{\lambda}-1}~~~. \eqno(11)$$
Now putting this expression into eq(8) and evaluating
the autocorrelation function one gets $\gamma=1/y_{\lambda}$.

From the scaling behavior of the spectrum in eq(9) one obtains for the finite
size corrections to the largest eigenvalues:
$$\lambda_0-\lambda_i(L) \sim L^{-y_{\lambda}}~~~, \eqno(12)$$
thus from eqs(6) and (12) the longitudinal correlation length
is $\xi_{\parallel} \sim L^{y_{\lambda}}$. In a finite system the
correlation length perpendicular to the (1,1) surface is limited by
the width of the strip
$\xi_{\perp} \sim L$, therefore the interface wandering
exponent in eq(7), which can be alternatively defined as
$\xi_{\perp} \sim \xi_{\parallel}^w$, is given by:
$$w=1/y_{\lambda}~~~. \eqno(13)$$
Thus indeed $w=\gamma$, as expected from scaling considerations.

In the following we calculate the interface fluctuations on the HK lattice
by two methods. First we study numerically the spectrum of the transfer
matrix, verify the validity of the scaling hypothesis and determine
the interfacial tension and the wandering exponent. Then
we apply an exact renormalization group transformation
and calculate analytical expressions for the critical exponents. 
\bigskip
{\bf III. Numerical Study of the Transfer Matrix}
\vskip 1cm
The transfer matrix of the interface problem in eq(4) is tridiagonal and could
be diagonalised by powerful methods[25]. In the specific problem, however, due
to the hierarchical structure of the
transfer matrix one can implement a very fast algorithm to calculate the
roots of the corresponding determinant.

We consider a finite system of size $L=2^l$ and express the corresponding
determinant
$D(2^l)$ by two subdeterminants of sizes $2^{l-1}$ and $2^{l-1}-1$, respectively, in the form:
$$D(2^l)=D(2^{l-1})D(2^{l-1})-D(2^{l-1}-1)D(2^{l-1}-1) \epsilon_{l-1}^2~~~.
\eqno(14a)$$
The symmetric determinant $\tilde D(2^l-2)$
of size $2^l-2$,
which is obtained from $D(2^l)$ by leaving out the first and last rows and columns,
can be similarly expressed as:
$$\tilde D(2^l-2)=D(2^{l-1}-1)D(2^{l-1}-1)-\tilde D(2^{l-1}-2) \tilde D(2^{l-1}-2)
\epsilon_{l-1}^2~~~. \eqno(14b)$$
Finally:
$$D(2^l-1)=D(2^{l-1})D(2^{l-1}-1)-D(2^{l-1}-1) \tilde D(2^{l-1}-2)
\epsilon_{l-1}^2~~~.
\eqno(14c)$$
These relations supplemented with $D(1)=-\lambda$, $D(2)=\lambda^2-\epsilon_0^2$
and $\tilde D(2)=\lambda^2-\epsilon_1^2$ define a fast procedure to
calculate the value of the determinant for very large sizes. For example
we could treat with this method slightly perturbed systems with $R \approx 1$ up to
sizes $L=2^{30}-2^{40}$.

The largest
eigenvalues calculated by this method all have the same type of finite size
dependence, thus the scaling hypothesis in Sec.II is indeed satisfied. The
leading eigenvalues calculated on the largest finite lattices are accurate
at least up to 10-12 digits. The gap exponents, describing the finite size
dependence of $\lambda_i(L)$ in eq(12), however could be obtained from the
raw data with a comperatively smaller accuracy, up to 5 digits. In
this case to increase accuracy we used sequence extrapolation methods, such
as the van den Broeck and Schwartz and the Bulirsch and Stoer methods[26]. 

The leading eigenvalue of the transfer matrix, which is connected to the
interfacial tension in eq(5) and the extrapolated values of the
the interface wandering exponent are listed on Table 1. One can
see that both the leading eigenvalue and the wandering exponent are
monotonically decreasing
as $R$ goes from one to zero. In the limit $R \to 0$ the interfacial tension in
eq(5) together with the wandering exponent go to zero, which is due to the fact
that the system tends to be separated into disconnected parts. More interesting
is the behavior of the wandering exponent around the homogeneous lattice point. As the
value of $R$ is lowered below one the wandering exponent jumps by a finite amount
of $\Delta w=0.0432799$ from $w=1/2$. In the renormalization group language
such type of behavior
corresponds to a relevant perturbation, which brings the system into
another stable fixed point. In the next section we shall explicitly construct
the RG transformation and determine exactly the wandering exponent.
\bigskip
{\bf IV. Renormalization Group Calculation}
\vskip 1cm
We are going to study the scaling behavior of the largest eigenvalues of the
transfer matrix in eq(4), which satisfy the second order difference equation:
$$0=T_{i,i+1} \psi_{i+1} - \lambda \psi_{i} + T_{i-1,i} \psi_{i-1}~~~, \eqno(15)$$
where in the thermodynamic limit the boundary terms are omitted. The structure
of the couplings which are connected to $T_{i,i+1}$ in eq(4) are shown on Fig.1. To
construct an exact
recursion we decimate out those sites, which are connected to a $\kappa_1$
coupling or equivalently to an $\epsilon_1$
matrix-element (denoted by crosses on Fig.1). We note, that the
same type of decimation was used by Maritan and Stella in their study of the
diffusion problem on the HK lattice[10]. One can see that after a decimation step the
$(\epsilon_0,\epsilon_1,\epsilon_0)$ triplet
will play the role of the renormalized $\epsilon_0'$, whereas the
other couplings will renormalize as $\epsilon_n'=\epsilon_{n+1}$, keeping the
value of $R$ and together the structure of the transfer matrix unchanged.

Performing the RG transformation first we denote the two neighbouring
sites to be decimated out by $i$ and $i+1$ and express $\psi_{i}$ and
$\psi_{i+1}$ as:
$$\psi_{i}=A \psi_{i-1} + B \psi_{i+2}$$
$$\psi_{i+1}=B \psi_{i-1} + A \psi_{i+2}~~~,\eqno(16)$$
where $A=\epsilon_0\lambda/(\lambda^2-\epsilon_1^2)$ and
$B=\epsilon_0\epsilon_1/(\lambda^2-\epsilon_1^2)$. Then the difference equations
in terms of the remaining, non-decimated spins have the same form as in eq(15),
provided the eigenvalue and the couplings transforms as:
$$\lambda'={\lambda-A\epsilon_0 \over B}$$
$$\epsilon_n'={\epsilon_{n+1} \over B}~~~,~~~n=1,2,...\eqno(17)$$
and $\epsilon_0'=\epsilon_0$. Thus the ratio of the sequence remains invariant
$R'=R$, as expected. As a consequence in the RG transformation
besides $\lambda$ only one coupling, say $\epsilon_1$ is enough to be
considered and the RG
transformation can be written as a two-parameter recursion:
$$\lambda'=\lambda {\lambda^2-\epsilon_1^2-\epsilon_0^2 \over \epsilon_0\epsilon_1}$$
$$\epsilon_1'=R {\lambda^2-\epsilon_1^2 \over \epsilon_0}~~~,\eqno(18)$$
where $\epsilon_0$ is the input value of the largest matrix-element.

The physically relevant fixed point of the transformation with $\lambda>0$
is given by:
$$\left( {\epsilon_1 \over \epsilon_0} \right)^*={R \over 1-R}~~~,
~~~\left( {\lambda \over \epsilon_0} \right)^*={\sqrt{1-R+R^2}\over 1-R}~~~,
\eqno(19)$$
which is stable for $0<R<1$. The eigenvalues of the linearised fixed point
transformation are roots of a quadratic equation and they are given as:
$$\Lambda_{1,2}={1 \over R} +R + {1 \over 2} \pm \left[ \left({1 \over R} +R + {1 \over 2}
\right)^2-2\right]^{1/2}~~~.\eqno(20)$$
The leading eigenvalue $\Lambda_1>1$ determines the scaling behavior
of the spectrum of the transfer matrix and the $y_{\lambda}$ scaling
dimension is given by:
$$y_{\lambda}={\log \Lambda_1 \over \log 2}~~~.\eqno(21)$$
The second eigenvalue of the RG transformation is $\Lambda_2<1$ and the
corresponding scaling field is irrelevant, thus the fixed point in eq(21) is
attractive and governs the critical properties of the physical model with
$\epsilon_1=R \epsilon_0$. It is seen from eq(19) that the fixed point of anomalous
interface fluctuations does not exist at the homogeneous point $R=1$, where the
fluctuations are characterised by the normal wandering exponent $w=1/2$.
Comparing the analytical results for $w=1/y_{\lambda}$ with those obtained by
finite size calculations in Table 1, we can say that the numerical results are
indeed very accurate they correspond to that in eq(21) at least up to six digits.
\bigskip
{\bf V. Discussion}
\vskip 1cm
In this paper we studied interface fluctuations on a layered
hierarchical lattice. The perturbation caused by inhomogeneous couplings
is relevant according to a
linear stability analysis and the observed interface fluctuations are indeed
anomalous. The
wandering exponent $w$ is a monotonically decreasing function of $R$ and
discontinuous at $R=1$. The fact that $w(R)<1/2$ can be understood, since
the interface preferentially stays on $\kappa_0$ lines and the probability to
visit a $\kappa_n$ line is rapidly decreasing with $n$. Consequently the
interface fluctuations are damped by the inhomogeneously distributed couplings.

One can estimate $w(R)$ in the limit $R \to 0$, when the probability of
a large interface fluctuation of height $h=2^n$ is primary
given by $p_n \sim \epsilon_n$, i.e. by the probability to have one step on the
$\kappa_n$ line. For such a fluctuation
the interface approximately takes $x \sim p_n^{-1} \sim R^{-n}$ steps,
thus the wandering
exponent in leading order is: $w(R)=-\log 2/ \log R$, which corresponds
to the aymptotic behavior of the analytical result in eq(22). We note
that in the $R \to 0$ limit the interface fluctuations can be described by
a Markovian process and then our problem is equivalent to the diffusion
of a particle in a hierarchical lattice, as studied in Ref[7-11].  

The HK-sequence used in this paper can be generalized, by having a general
$\nu$-ary character[27],
instead of $\nu=2$ used in eq(2). Then one has in eq(2) $h=R^n(\nu m + \mu)$,
with $\mu=1,2,...,\nu-1$. According to our numerical and analytical
investigations for $\nu=3$ and $\nu=4$ the main characteristics of
interface fluctuations remain the same as for $\nu=2$: the
wandering exponent has a jump at $R=1$ and varies with $R$. For $\nu=3$ we
obtained the analytical result:
$$w_{\nu=3}= {\log 3 \over \log \Lambda_{\nu=3}}$$
$$\Lambda_{\nu=3}=2 \left( {1 \over R} +R + 1 \right)+\left[4 \left( {1 \over R}
 +R + 1 \right)^2-3 \right]^{1/2}~~~.\eqno(22)$$
 
As mentioned before the problem studied in this paper is related to the
diffusion process on hierarchical lattices[11].
Our problem, however, can be formulated as the
quantum-mechanical diffusion process of a particle which is represented
by a wave
packet and placed on a one-dimensional HK-potential. Then $x$ and $h(x)$
correspond to the $t$ time and the position of the particle at the given
time step, respectively, while
the transfer matrix describes time evaluation. According to our results in
a one-dimensional hierarchical potential the width of the wave
packet will grow in time anomalously as $t^{w(R)}$.

Our final remark concerns some similarities of our results to that of
interface fluctuations in a repulsive, inhomogeneous surface potential,
decaying as $\sim l^{-\omega}$, where $l$ measures the distance from the
surface[28]. In two-dimensions for $\omega<2$ the perturbation is relevant
and the interface wandering exponent takes the anomalous value: $w=1/\omega
>1/2$[29]. In this problem, however, the perturbation is confined to the surface,
furthermore the wandering exponent is continuous at $\omega=2$.
\vskip 1cm
Acknowledgement: F.I. is indebted to L. Turban for valuable comments and for
hospitality in Nancy. He also acknowledges useful discussions with P.
Sz\'epfalusy and A. S\"ut\H o. This
work has been supported by an exchange program of the CNRS-Hungarian Academy
of Sciences and by the Hungarian National Research Fund under grant numbers:
OTKA TO12830 and OTKA TO17485. The Laboratoire de Physique du Solide is Unit\'e
de Recherche Associ\'ee au C.N.R.S. No 155.

\vfill
\eject
{\bf References}
\bigskip
\item{ [1]} H.A. Simon and A. Ando, Econometrica 29, 111 (1961)
\bigskip
\item{ [2]} J.A. Feigenbaum and P.G.A. Freund, Preprint EFI-95-58.
\bigskip
\item{ [3]} H. Saleur, C.G. Sammis and D. Sornette, Preprint USC-95-02
\bigskip
\item{ [4]} R.H. Austin, K.W. Berson, L. Eisenstein, L.H. Frauenfelder
and I.C. Gunsalus, Bio\-chem. 14, 5355 (1975)
\bigskip
\item{ [5]} M. M\'ezard, G. Parisi, N. Sourlas, G. Toulouse and M. Virasoro,
Phys. Rev. Lett. 52, 1156 (1984)
\bigskip
\item{ [6]} B.A. Huberman and T. Hogg, Phys. Rev. Lett. 52, 1048 (1984)
\bigskip
\item{ [7]} B.A. Huberman and M. Kerszberg, J. Phys. A18, L331 (1985)
\bigskip
\item{ [8]} S. Teitel, D. Kutasov and E. Domany, Phys. Rev. B36, 684 (1987)
\bigskip
\item{ [9]} M. Schreckenberg, Z. Phys. B60, 483 (1985); A.T. Ogielski and
D.L. Stein, Phys. Rev. Lett. 55, 1634 (1985); G. Paladin, M. M\'ezard and
C. de Dominicis, J. Phys. Lett. 46, L985 (1985)
\bigskip
\item{[10]} A. Maritan and A.L. Stella, J. Phys. A19, L269 (1986)
\bigskip
\item{[11]} A. Giacomtti, A. Maritan and A.L. Stella, Int. J. Mod. Phys. B5, 709 (1991)
\bigskip
\item{[12]} J.M. Luck, J. Stat. Phys. 72, 417 (1993); F. Igl\'oi, J. Phys. A26, L703 (1993);
J.M. Luck, Europhys. Lett. 24, 359 (1993)
\bigskip
\item{[13]} A.B. Harris, J. Phys. C7, 1671 (1974)
\bigskip
\item{[14]} M. Queff\'elec, {\it Substitutional Dynamical
Systems-Spectral Analysis}, Lecture Notes in Mathemathics, Vol 1294
ed. A. Dold and B. Eckmann (Springer, Berlin, 1987)
\bigskip
\item{[15]} C.A. Tracy, J. Phys. A21, L603 (1988); F. Igl\'oi, J. Phys. A21, L911 (1988);
 G.V. Benza, Europhys. Lett. 8, 321 (1989)
\bigskip
\item{[16]} L. Turban, F. Igl\'oi and B. Berche, Phys. Rev. B49, 12695
(1994); F. Igl\'oi and L. Turban, Europhys. Lett. 27, 91 (1994);
L. Turban, P-E. Berche and B. Berche, J. Phys. A27,
6349 (1994)
\bigskip
\item{[17]} B. Berche, P-E. Berche, M. Henkel, F. Igl\'oi, P.
Lajk\'o, S. Morgan and L. Turban, J. Phys. A28, L165 (1995); P-E. Berche, B. Berche
and L. Turban (unpublished); F. Igl\'oi and P. Lajk\'o (unpublished)
\bigskip
\item{[18]} A.L. Stella, M.R. Swift, J.G. Amar, T.L. Einstein, M.W.
Cole and J.R. Banavar, Phys. Rev. Lett. 23, 3818 (1993)
\bigskip
\item{[19]} Z. Lin and M. Goda, Phys. Rev. B51, 6093 (1995)
\bigskip
\item{[20]} F. Igl\'oi, P. Lajk\'o and F. Szalma, Phys. Rev. B52, 7159 (1995)
\bigskip
\item{[21]} C.L. Henley and R. Lipowsky, Phys. Rev. Lett. 59, 1679 (1987)
\bigskip
\item{[22]} M.E. Fisher, J. Chem. Soc. Faraday Trans. 82, 1569 (1986)
\bigskip
\item{[23]} T.W. Burkhardt, J. Phys. A14, L63 (1981)
\bigskip
\item{[24]} V. Privman and N.M. \v Svraki\'c,{\it Directed Models of Polymers,
Interfaces and Finite-Size Properties}, in Lecture Notes in Physics, 338 (Springer, Berlin, 1989)
\bigskip
\item{[25]} C. Lanczos, J. Res. Nat. Bur. Stand. 45, 255 (1950)
\bigskip
\item{[26]} M. Henkel and G. Sch\"utz, J. Phys. A21, 2617 (1988)
\bigskip
\item{[27]} W.P. Keirstead and B. Huberman, Phys. Rev. A36, 5392 (1987)
\bigskip
\item{[28]} R. Lipowsky and T.M. Nieuwenhuizen, J. Phys. A21, L89 (1988)
\bigskip
\item{[29]} F. Igl\'oi, Europhys. Lett. 19, 305 (1992)
\vfill
\eject
$$\vbox{\settabs 3 \columns
\+$R$&$\lambda_0/\epsilon_0$&$w=1/y_{\lambda}$\cr
\+&&\cr
\+1.&2.&0.5\cr
\+0.999&1.99800894&0.4567199\cr
\+0.9&1.82853274&0.4551092\cr
\+0.75&1.62218648&0.4451438\cr
\+0.5&1.35286081&0.4004540\cr
\+0.25&1.14948652&0.3110577\cr
\+0.1&1.05381456&0.2272971\cr
\+0.001&1.00050038&0.0911867\cr
\+&&\cr}$$
\vskip 1cm
\centerline{Table 1}

The leading eigenvalue and the
corresponding interface fluctuation exponent from numerical diagonalization
of the transfer matrix for different values of the
hierarchical parameter.
\eject
{\bf Figure captions}
\vskip 1cm
\item{Fig.1:} Structurless interface on a diagonally layered square lattice.
The values of the couplings, which follow the hierarchical HK sequence in eq(2)
are indicated below. Sites to be decimated out in the RG transformation are
marked by $X$.
\vfill
\eject
\end